\documentclass{iaus1}
\usepackage{graphicx}

\title[First Middle East-Africa IAU Regional Meeting] 
{Star Formation in Various Environments}

\author[N. Brosch, E. Almoznino, O. Spector and A. Zitrin]   
{Noah Brosch$^1$, Elhanan Almoznino$^1$, Oded Spector$^1$, Adi Zitrin$^1$}

\affiliation{$^1$The Wise Observatory and the Beverly and Raymond Sackler School
of Physics and Astronomy, \\ Tel Aviv University, Tel Aviv 69978, Israel \\
email: {\tt noah@wise.tau.ac.il}}

\pubyear{2008}
\volume{xxx}  
\pagerange{119--126}
 \jname{First Middle East-Africa IAU Regional
Meeting} \editors{A.C. Editor, B.D. Editor \& C.E. Editor, eds.}
\begin{document}

\maketitle

\begin{abstract}
We describe studies of star formation in various galaxies using primarily observations from the Wise Observatory. In addition to surface photometry in the broad band UBVRI, we also use a set of narrow-band H$\alpha$ filters tuned to different redshifts to isolate the emission line. With these observational data, and using models of evolutionary stellar populations, we unravel the star formation histories of the galaxies and connect them to other parameters, such as the galaxy environment.

\end{abstract}

\firstsection 

\vspace{5mm}

The process of star formation is basic in the formation and evolution of galaxies on all scales of the Universe. However, despite the decades of research, this process is still not fully understood. Many factors combine to trigger star formation (SF) once the available ingredients (primarily interstellar gas) are present. These are of different kinds, such as spiral density waves (SDWs) in large spiral galaxies, shock-wave triggered SF in objects where massive stars already formed and evolved, etc.

The basic research question is ``How do galaxies form stars?''. In order to reach an answer one needs to isolate, as much as possible, the different factors that affect SF. In particular, one should select galaxies where the large-scale SF triggers, such as SDWs, do not operate. If one uses dwarf galaxies (DGs) the influence of such global SF triggers can be minimized. For this reason, we initiated at the Wise Observatory a program to study SF in dwarf galaxies in an attempt to clarify what are the important factors deciding when a galaxy is forming stars and when it is not.

We checked for influence of neighborhood by sample selection. Our first attempts were with DGs in the Virgo Cluster (VC), to identify factors affecting SF in a relatively dense and nearby galaxy environment where DGs of all kinds can be easily observed. We studied a sample of blue compact DGs (BCDs) that had previous 21-cm line measurements, which was characterized as either high or low HI content (Almoznino \& Brosch 1998a, b) as an example of galaxies with bright and concentrated young stellar populations, and another sample of DGs in the VC that showed low surface brightness levels (LSBs) and were selected in a  similar manner regarding their HI content (Heller et al. 1999  a,b). Both samples were studied with CCD imaging in broad bands (UBVRI) and in rest-frame H$\alpha$ using 50\AA\, wide filters.

We compared the photometry results with evolutionary synthesis models including either short
SF bursts or continuous SF. We found that most LSBs and BCDs in the VC sample formed stars in short bursts. More than one stellar generation per galaxy was observed. No obvious influences of interactions, even in a cluster environment, were detected for these samples.

In order to mitigate the influence of the environment, we selected a sample of blue compact galaxies located in voids. This sample was observed in a similar manner and the results are presently being analyzed (Zitrin et al., in preparation). Another aspect of this project was a search for H$\alpha$ emitting knots in the immediate vicinity of DGs in voids that showed signs of star formation (Brosch et al. 2006). This revealed a large number of small, compact, and faint knots emitting the H$\alpha$ line at the same redshift as that of the DG around which the search was conducted, emphasizing that locations with well-detected SF will support SF at other nearby sites as well, even if located in cosmic voids.

We are now studying a sample of very isolated galaxies, selected to be more than 3h$^{-1}$ Mpc away from any other galaxy (Spector et al., in preparation). The degree of isolation is tested optically and with HI measurements from the ALFALFA survey, to eliminate cases of neighbouring extreme LSBs (with measurable neutral gas) that could affect the SF properties of the target galaxies. These extremely isolated galaxies have had the least chance of interaction with another galaxy that could influence SF for a Hubble time. In this case as well we study SF with UBVRI and rest-frame H$\alpha$ CCD imaging.

The preliminary results obtained for the sample galaxy VIII Zw 202 show that there also the SF probably took place in (at least) two episodes of short SF bursts. The study of this sample is continuing.

We also identified an isolated group of DGs in the nearby Universe within $\leq$7 Mpc (Zitrin et al., in preparation). The group consists of 14 DGs [M(B)$\geq$--18], most with HI measurements from the ALFALFA survey. The HI line is, in all cases, narrow to very narrow, indicating low masses. The linear galaxy configuration, with most objects on an almost straight line projected on the sky that is $\sim$1Mpc, hints at a possible common location in 3D space for all objects and a possible common origin.

WO observations detected H$\alpha$ emission in all galaxies, indicating present-day star formation. Comparison with evolutionary population synthesis models indicates that all galaxies had a SF burst a few Myrs ago. The results beg the following question: ``Why do we observe synchronized SF in a linear group of galaxies, where no obvious signs of interaction with a major nearby galaxy are present?''

At this point we can only present a conjuncture: SF occurs because of intergalactic gas collapsing onto a DM filament. The DGs we observe in the linear filament are destined to form a major galaxy in the future by merging, as predicted by the hierarchical structure formation scenarios. A possible alternative could be interaction-triggered SF, but we stress that no interaction signs are observed! If our conjuncture is correct, this represents the nearest galaxy collection where we witness hierarchical clustering.

{\bf Preliminary conclusions:}

\begin{itemize}

\item Neighborhood influence on star formation of non-cluster galaxies are not significant, even in environments as these as the VC.

\item Interactions as star formation triggers may be important in much denser environments (cores of clusters, compact groups).

\item Our results stress the need to understand the dynamics of dwarf galaxies; how much is Dark Matter and how does this Dark Matter influence the star formation?

\end{itemize}





\begin{thebibliography}{}

\bibitem[Almoznino
\& Brosch(1998b)]{1998MNRAS.298..931A} Almoznino, E., \& Brosch, N.\ 1998b, MNRAS, 298, 931


\bibitem[Almoznino
\& Brosch(1998a)]{1998MNRAS.298..920A} Almoznino, E., \& Brosch, N.\ 1998a, MNRAS, 298, 920


\bibitem[Brosch
\& Almoznino(1997)]{1997ApL&C..36..303B} Brosch, N., \& Almoznino, E.\ 1997, Astrophysical Letters Communications, 36, 303

\bibitem[Brosch et al.(2006)]{2006MNRAS.368..864B} Brosch, N., Bar-Or, C.,
\& Malka, D.\ 2006, MNRAS, 368, 864

\bibitem[Heller et al.(1999a)]{1999MNRAS.304....8H} Heller, A., Almoznino,
E., \& Brosch, N.\ 1999a, MNRAS, 304, 8


\bibitem[Heller et al.(1999b)]{1999ASPC..170..282H} Heller, A., Almoznino,
E., \& Brosch, N.\ 1999b, The Low Surface Brightness Universe, 170, 282

\end{thebibliography}
\end{document}